\documentclass[prx,aps,twocolumn,eqsecnum,superscriptaddress,showpacs,amsmath]{revtex4-2}
\usepackage{amsmath,amssymb,graphicx,graphics,color}
\usepackage{dcolumn}
\usepackage{bm}
\usepackage{psfrag}
\usepackage{graphicx}
\usepackage{latexsym}
\usepackage{epsfig}
\usepackage{color}
\usepackage{amsmath}
\usepackage{subfigure}
\usepackage{multirow}
\usepackage{comment}
\usepackage{braket}

\begin{document}

\title{Competing spin-orbital singlet states in the 4$d^4$ honeycomb ruthenate Ag$_3$LiRu$_2$O$_6$}
\author{T. Takayama}
\affiliation{Max Planck Institute for Solid State Research, Heisenbergstrasse
  1, 70569 Stuttgart, Germany}
\affiliation{Institute for Functional Matter and Quantum Technologies, University of Stuttgart, Pfaffenwaldring 57, 70550 Stuttgart, Germany}
\author{M. Blankenhorn}
\affiliation{Max Planck Institute for Solid State Research, Heisenbergstrasse
  1, 70569 Stuttgart, Germany}
\affiliation{Institute for Functional Matter and Quantum Technologies, University of Stuttgart, Pfaffenwaldring 57, 70550 Stuttgart, Germany}
\author{J. Bertinshaw}
\affiliation{Max Planck Institute for Solid State Research, Heisenbergstrasse
  1, 70569 Stuttgart, Germany}
\author{D. Haskel}
\affiliation{Advanced Photon Source, Argonne National Laboratory, Argonne, Illinois 60439, USA}
\author{N.A. Bogdanov}
\affiliation{Max Planck Institute for Solid State Research, Heisenbergstrasse
  1, 70569 Stuttgart, Germany}
\author{K. Kitagawa}
\affiliation{Department of Physics, University of Tokyo, 7-3-1 Hongo, Bunkyo-ku, Tokyo 113-0033, Japan}
\author{A. N. Yaresko}
\affiliation{Max Planck Institute for Solid State Research, Heisenbergstrasse
  1, 70569 Stuttgart, Germany}
\author{A. Krajewska}
\affiliation{Max Planck Institute for Solid State Research, Heisenbergstrasse
  1, 70569 Stuttgart, Germany}
\affiliation{Institute for Functional Matter and Quantum Technologies, University of Stuttgart, Pfaffenwaldring 57, 70550 Stuttgart, Germany}
\affiliation{ISIS Neutron and Muon Source, STFC Rutherford Appleton Laboratory, Chilton, Didcot, Oxon OX11 0QX, UK}
\author{S. Bette}
\affiliation{Max Planck Institute for Solid State Research, Heisenbergstrasse
  1, 70569 Stuttgart, Germany}
\author{G. McNally}
\affiliation{Max Planck Institute for Solid State Research, Heisenbergstrasse
  1, 70569 Stuttgart, Germany}
\author{A. S. Gibbs} 
\affiliation{ISIS Neutron and Muon Source, STFC Rutherford Appleton Laboratory, Chilton, Didcot, Oxon OX11 0QX, UK}
\affiliation{School of Chemistry, University of St Andrews, St Andrews, KY16 9ST, UK}
\author{Y. Matsumoto}
\affiliation{Max Planck Institute for Solid State Research, Heisenbergstrasse
  1, 70569 Stuttgart, Germany}
\author{D. P. Sari}
\affiliation{Graduate School of Engineering and Science, Shibaura Institute of Technology, 307 Fukasaku, Minuma, Saitama 337-8570, Japan}
\author{I. Watanabe}
\affiliation{Advanced Meson Science Laboratory, RIKEN Nishina Center for Accelerator-Based Science, Wako, Saitama 351-0198, Japan}
\author{G. Fabbris}
\affiliation{Advanced Photon Source, Argonne National Laboratory, Argonne, Illinois 60439, USA}
\author{W. Bi}
\affiliation{Advanced Photon Source, Argonne National Laboratory, Argonne, Illinois 60439, USA}
\affiliation{Department of Physics, University of Alabama at Birmingham, Birmingham, Alabama 35294, USA}
\author{T. I. Larkin}
\affiliation{Max Planck Institute for Solid State Research, Heisenbergstrasse
  1, 70569 Stuttgart, Germany}
\author{K. S. Rabinovich}
\affiliation{Max Planck Institute for Solid State Research, Heisenbergstrasse
  1, 70569 Stuttgart, Germany}
\author{A. V. Boris}
\affiliation{Max Planck Institute for Solid State Research, Heisenbergstrasse
  1, 70569 Stuttgart, Germany}
\author{H. Ishii}
\affiliation{National Synchrotron Radiation Research Center, Hsinchu 30076, Taiwan}
\author{H. Yamaoka}
\affiliation{RIKEN SPring-8 Center, Sayo, Hyogo 679-5148, Japan}
\author{T. Irifune}
\affiliation{Geodynamics Research Center, Ehime University, Matsuyama, 790-8577, Japan}
\author{R. Bewley} 
\affiliation{ISIS Neutron and Muon Source, STFC Rutherford Appleton Laboratory, Chilton, Didcot, Oxon OX11 0QX, UK}
\author{C. J. Ridley} 
\affiliation{ISIS Neutron and Muon Source, STFC Rutherford Appleton Laboratory, Chilton, Didcot, Oxon OX11 0QX, UK}
\author{C. L. Bull} 
\affiliation{ISIS Neutron and Muon Source, STFC Rutherford Appleton Laboratory, Chilton, Didcot, Oxon OX11 0QX, UK}
\affiliation{School of Chemistry, University of Edinburgh, David Brewster Road, Edinburgh EH9 3FJ, UK}
\author{R. Dinnebier}
\affiliation{Max Planck Institute for Solid State Research, Heisenbergstrasse
  1, 70569 Stuttgart, Germany}
\author{B. Keimer}
\affiliation{Max Planck Institute for Solid State Research, Heisenbergstrasse
  1, 70569 Stuttgart, Germany}
\author{H. Takagi}
\affiliation{Max Planck Institute for Solid State Research, Heisenbergstrasse
  1, 70569 Stuttgart, Germany}
\affiliation{Institute for Functional Matter and Quantum Technologies, University of Stuttgart, Pfaffenwaldring 57, 70550 Stuttgart, Germany}
\affiliation{Department of Physics, University of Tokyo, 7-3-1 Hongo, Bunkyo-ku, Tokyo 113-0033, Japan}

\date{\today}
\begin{abstract}
When spin-orbit-entangled $d$-electrons reside on a honeycomb lattice, rich quantum states are anticipated to emerge, as exemplified by the $d^5$ Kitaev materials. Distinct yet equally intriguing physics may be realized with a $d$-electron count other than $d^5$. We found that the layered ruthenate Ag$_3$LiRu$_2$O$_6$ with $d^4$ Ru$^{4+}$ ions at ambient pressure forms a honeycomb lattice of spin-orbit-entangled singlets, which is a playground for frustrated excitonic magnetism. Under pressure, the singlet state does not develop the expected excitonic magnetism but experiences two successive transitions to other nonmagnetic phases, first to an intermediate phase with moderate distortion of honeycomb lattice, and eventually to a high-pressure phase with very short Ru-Ru dimer bonds. While the strong dimerization in the high-pressure phase originates from a molecular orbital formation as in the sister compound Li$_2$RuO$_3$, the intermediate phase represents a spin-orbit-coupled $J$-dimer state which is stabilized by the admixture of upper-lying $J_{\rm eff} = 1$-derived states. We argue that the $J$-dimer state is induced by a pseudo-Jahn-Teller effect associated with the low-lying spin-orbital excited states and is unique to spin-orbit-entangled $d^4$ systems. The discovery of competing singlet phases demonstrates rich spin-orbital physics of $d^4$ honeycomb compounds and paves the way for realization of unconventional magnetism.
\end{abstract}


\maketitle
\section{Introduction}

The interplay of electron correlation and spin-orbit coupling (SOC) in 4$d$ and 5$d$ transition-metal compounds has been recognized as a key ingredient in realizing unprecedented electronic phases. The energy scale of SOC for 4$d$ and 5$d$ transition metal ions is comparable to the other relevant electronic parameters such as Hund’s coupling, inter-site hopping, and non-cubic crystal field in those systems, giving rise to the formation of a spin-orbit-entangled wave function \cite{Abragam70, Takayama2021}. Magnetic couplings between spin-orbit-entangled moments are often anisotropic reflecting the inherited orbital degree of freedom, which is distinct from those of spin-only magnetic moments and realizes rich and exotic magnetic ground states \cite{Takayama2021, Khaliullin2005}. Serving as a prime example are compounds with a honeycomb-lattice of spin-orbit-entangled moments. Those with $d^5$ and $d^1$ ions are proposed theoretically to host a novel class of quantum liquids, the Kitaev spin liquid \cite{Jackeli2009,Takagi2019} and a SU(4) spin-orbital liquid \cite{Yamada2018}, respectively. 

Honeycomb-lattice compounds with $d^4$ transition-metal ions, which are coordinated octahedrally with ligand anions, are another playground for exotic magnetism. All the four $d$ electrons are accommodated in the low-lying $t_{2g}$ manifold because of the large cubic crystal field for 4$d$ and 5$d$ transition metal ions. Hund’s coupling leads to the formation of spin moment $S$ = 1 and effective orbital moment $L_{\rm eff}$ = 1 in the $LS$-coupling scheme. SOC splits $S$ = 1 $L_{\rm eff}$ = 1 manifold into spin-orbit-entangled $J_{\rm eff}$ = 0, 1 and 2 states. The ground state for an isolated ion is the nonmagnetic $J_{\rm eff}$ = 0 singlet \cite{Abragam70}. It is proposed theoretically that a $J_{\rm eff}$ = 0 Mott insulator experiences a quantum phase transition to a magnetically ordered state when the exchange interactions through the upper-lying $J_{\rm eff}$ = 1 state overcome the excitation gap from the $J_{\rm eff}$ = 0 to $J_{\rm eff}$ = 1 states \cite{Khaliullin2013, Meetei2015}. This magnetic transition can be viewed as a condensation of $J_{\rm eff}$ = 1 triplet excitons and is dubbed excitonic magnetism. The antiferromagnetically ordered state of the layered perovskite Ca$_2$RuO$_4$ was recently shown to be a realization of such excitonic magnetism and a Higgs-mode excitation was identified in the spin-wave dispersion measured by inelastic neutron scattering and in Raman spectra \cite{Jain2017, Souliou2017}.

The excitonic magnetism of Ca$_2$RuO$_4$ with the corner-shared RuO$_6$ octahedra is produced predominantly by Heisenberg-type exchange interactions through the excited $J_{\rm eff}$ = 1 states in the nearly 180$^{\circ}$ Ru-O-Ru bonds. In the network of edge-shared octahedra with 90$^{\circ}$ bonding geometry, the exchange interaction via the $J_{\rm eff}$ = 1 triplet is dependent on the dominant hopping process and anisotropic in contrast to the corner-shared case. The anisotropic coupling brings about a frustration in developing excitonic magnetism, especially on a honeycomb lattice. When the hopping via anion $p$-orbitals, $d$-$p$-$d$ hopping, is dominant in the 90$^{\circ}$ bond, the exchange interaction takes the form of bond-dependent XY-interactions. Magnetic correlations in such a case develop only along the one-dimensional zigzag chain segments of honeycomb lattice, which is predicted to give rise to a spin-nematic state \cite{Khaliullin2013}. If the direct $d$-$d$ hopping across the edges is dominant, a bond-dependent Ising interaction of the Kitaev-type is proposed to appear \cite{Anisimov2019, Chaloupka2019}. Strong frustration originating from the Kitaev-type interaction is expected to suppress the long-range magnetic ordering and lead to a $J_{\rm eff}$ = 1 triplon liquid phase, which is regarded as a bosonic analog of Kitaev honeycomb model. When both hopping processes contribute with comparable magnitudes, condensation of $J_{\rm eff} = 1$ triplet, namely long-range magnetic ordering with soft moments, takes place as in Ca$_2$RuO$_4$ \cite{Chaloupka2019}. With introducing lattice degree of freedom, even richer variety of spin-orbit-entangled phases may emerge. The $d^4$ honeycomb systems are therefore a promising playground to explore such exotic magnetic ground states.

There are $d^4$ honeycomb materials which are nonmagnetic. They, however, do not represent a good starting point to explore the expected exotic magnetism. The 5$d^4$ honeycomb iridate NaIrO$_3$, a Na-deficient analogue of Na$_2$IrO$_3$, displays almost temperature-independent magnetic susceptibility \cite{Wallace2015}. The nonmagnetic insulating state implies that NaIrO$_3$ is viewed as a $J_{\rm eff}$ = 0 Mott insulator if the $LS$-coupling scheme is justified. The large SOC-induced gap of ~0.4 eV to the $J_{\rm eff}$ = 1 triplet in 5$d^4$ iridates \cite{Kusch2018}, however, is hard to reach by the softening due to the exchange interactions, placing it far away from the excitonic magnetism. Besides, the large SOC is comparable to the Hund’s coupling and a $jj$-coupling character is non-negligible in NaIrO$_3$. The 4$d^4$ honeycomb systems with a moderate SOC should be more suitable for realizing excitonic and related magnetisms. The well-known 4$d^4$ honeycomb ruthenate Li$_2$RuO$_3$, however, experiences a strong Ru-Ru dimerization below $\sim$540 K accompanied with a formation of molecular orbitals \cite{Miura2007,Miura2009}. In the dimerized state, the spin and orbital degrees of freedom are fully quenched, hampering the formation of spin-orbit-entangled state.

In order to realize the $J_{\rm eff}$ = 0 state on a honeycomb lattice, strong Ru-Ru dimerization needs to be evaded. We utilized a soft-chemical reaction to suppress the Ru-Ru dimerization of Li$_2$RuO$_3$. The silver-intercalated Li$_2$RuO$_3$, Ag$_3$LiRu$_2$O$_6$, shows no dimerization down to the lowest temperature measured at ambient pressure \cite{Kimber2010}. The magnetic and spectroscopic measurements, together with the quantum chemistry calculation, indicate that Ag$_3$LiRu$_2$O$_6$ is a Mott insulator and hosts a spin-orbit-entangled singlet state derived from $J_{\rm eff}$ = 0 ($J$-singlet). With the application of pressure, Ag$_3$LiRu$_2$O$_6$ exhibits two successive phase transitions from the $J$-singlet state to other nonmagnetic phases, accompanied by structural distortions. The structure of the higher-pressure phase comprises very short Ru-Ru bonds as in Li$_2$RuO$_3$, indicating the strong Ru-Ru dimerization and hence a molecular-orbital formation. The intermediate phase in contrast experiences only a modest distortion of the honeycomb lattice. We argue that the intermediate phase hosts spin-orbit-coupled weak dimers where the lattice distortion leads to admixture of $J_{\rm eff} = 1$-derived states into the singlet state and lowers its energy. The lattice distortion is viewed as a pseudo-Jahn-Teller effect associated with low-lying spin-orbital excitations, which is potentially inherent in $d^4$ honeycomb compounds. The emergence of competing spin-orbital singlet phases points to the intricate interplay between SOC, exchange interactions and the lattice in the honeycomb lattice of $d^4$ ions, and the understanding of the phase competition should give a clue to realize exotic magnetic ground states.

\section{Results}

\subsection{Crystal structure and electronic structure of Ag$_3$LiRu$_2$O$_6$}

\begin{figure*}[t]
\begin{center}
\includegraphics[scale=0.15]{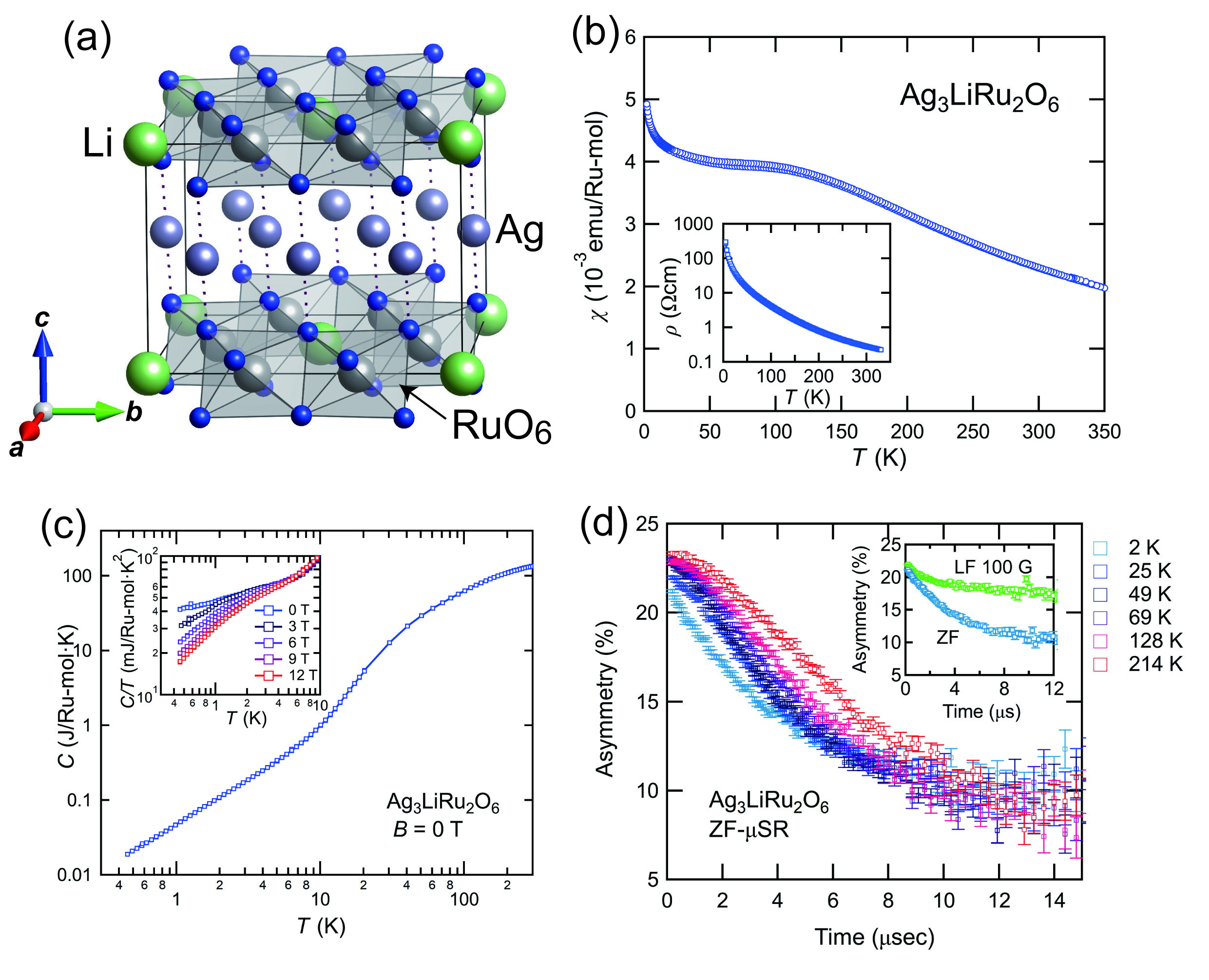}
\caption{Absence of magnetic order in the honeycomb ruthenate Ag$_3$LiRu$_2$O$_6$. (a) Crystal structure of Ag$_3$LiRu$_2$O$_6$. The RuO$_6$ octahedra compose edge-shared honeycomb layers and Li ions locate at the center of honeycomb. Ag ions form O-Ag-O dumbbell bonds between the layers. (b) Magnetic susceptibility $\chi$($T$) as a function of temperature. The inset shows the temperature-dependent resistivity. (c) Specific heat $C(T)$ at zero magnetic field. The inset shows the $C(T)$ divided by temperature at low temperatures under magnetic fields. The strong suppression of $C(T)/T$ at low temperatures by magnetic fields suggests that the low-temperature contributions originate from localized spin defects. (d) Zero-field (ZF) muon spin relaxation at various temperatures down to 2 K. The inset shows the longitudinal field (LF) measurement with an applied magnetic field of 100 Gauss at 5 K, together with the ZF data at 2 K.
}
\label{fig:1}
\end{center}
\end{figure*}

Ag$_3$LiRu$_2$O$_6$ was obtained by an ion-exchange reaction of Li$_2$RuO$_3$ and AgNO$_3$ \cite{Kimber2010} (see Supplemental Material for the experimental and theoretical methods \cite{SM}). The chemical formula of Li$_2$RuO$_3$ can be conveniently rewritten as 1/2 Li(I)$_3$Li(II)Ru$_2$O$_6$ where the Li(I) ions occupy the interlayer sites between the Ru-honeycomb layers and Li(II) ion resides at the center of Ru honeycomb lattice. Ag$_3$LiRu$_2$O$_6$ corresponds to the case that all the interlayer Li(I) ions of Li$_2$RuO$_3$ are replaced by Ag ions. The full replacement of Li(I) with Ag was confirmed by the structural refinement of powder x-ray diffraction \cite{Bette2019}. The intercalated Ag$^+$ ions form covalent O-Ag-O dumbbell bonds with oxygen ions in the Li(II)Ru$_2$O$_6$ layers above and below as in the delafossite oxides \cite{Sheets2008} (Fig.~\ref{fig:1}(a)). Li$_2$RuO$_3$ has three inequivalent Ru-Ru bonds on the honeycomb lattice at room temperature, which can be classified into two distinct groups, one short bond with a length of 2.567 \AA~ representing the dimer formation and the other two long bonds with similar lengths of 3.046 \AA~ and 3.049 \AA~ \cite{Miura2007}. In contrast, the Ru honeycomb lattice of Ag$_3$LiRu$_2$O$_6$ consists of only two inequivalent Ru-Ru bonds with similar lengths of 3.01(2) \AA~ and 3.019(10) \AA~ and is thus almost regular \cite{Bette2019}, indicating the absence of Ru-dimerization. We argue that the strong interlayer chemical bond, originating from O 2$p$ and Ag $d$ 3$z^2$-1 orbitals (Supplemental Fig. S1 \cite{SM}), prevents the distortion of the honeycomb lattice and hence suppresses the dimerization.

Ag$_3$LiRu$_2$O$_6$ was originally reported to be metallic, based on the reduced resistivity compared to that of Li$_2$RuO$_3$ \cite{Kimber2010}. However, we confirmed the semiconducting behavior of resistivity below room temperature (the inset of Fig.~\ref{fig:1}(b)) with an activation energy $E_{\rm a}$ $\sim$ 700 K at room temperature and the presence of a small charge gap of the order of $\sim$0.1 eV in the optical conductivity (Supplemental Figs. S2 and S3 \cite{SM}). The density functional theory (DFT) calculation yields a metallic ground state with a relatively high density of states (Fig. S1 \cite{SM}). We therefore argue that Ag$_3$LiRu$_2$O$_6$ is a weak Mott insulator produced by the moderate electron correlation of Ru 4$d$ electrons. The successful analysis of magnetic properties based on the localized 4$d^4$ electrons supports further the Mott insulating state of Ag$_3$LiRu$_2$O$_6$. The much smaller resistivity compared with that of Li$_2$RuO$_3$ may be attributed to the weak Mottness and the absence of large bonding-antibonding splitting of $d$-electron molecular orbitals. The presence of Ag $d$ 3$z^2$-1 – O 2$p$ antibonding states near the chemical potential may also contribute further to reduce the charge gap.

\subsection{Nonmagnetic ground state at ambient pressure}

The absence of strong Ru-Ru dimers in Ag$_3$LiRu$_2$O$_6$ appears to restore the magnetism of Ru 4$d$ electrons. Magnetic susceptibility $\chi$($T$) on cooling from room temperature first shows a Curie-like increase and then crossovers to a temperature-independent paramagnetic susceptibility below $\sim$100 K as displayed in Fig.~\ref{fig:1}(b). This behavior is reminiscent of van Vleck susceptibility observed in $J$ = 0 Eu$^{3+}$ compounds with $J$ = 1 and higher excitations \cite{Takikawa2010}, suggesting a spin-orbit-entangled singlet state. No signature of magnetic transition is indeed observed in $\chi$($T$), consistent with the singlet ground state.

The absence of magnetic order is corroborated indirectly by the specific heat $C(T)$ and directly by the NMR and muon spin relaxation ($\mu$SR) measurements. The temperature dependent $C(T)$ shown in Fig.~\ref{fig:1}(c) does not show any signature of phase transition below room temperature. There is a large $T$-linear $C(T)$ at low temperatures, which is suppressed drastically by applying a magnetic field roughly up to a temperature corresponding to the Zeeman energy. The magnetic field suppression scaled by the Zeeman energy indicates that it originates predominantly from almost free spin defects \cite{Kitagawa2018}, which is represented by the low-temperature Curie-tail superposed on the temperature-independent van Vleck susceptibility.

\begin{figure*}[t]
\begin{center}
\includegraphics[scale=0.15]{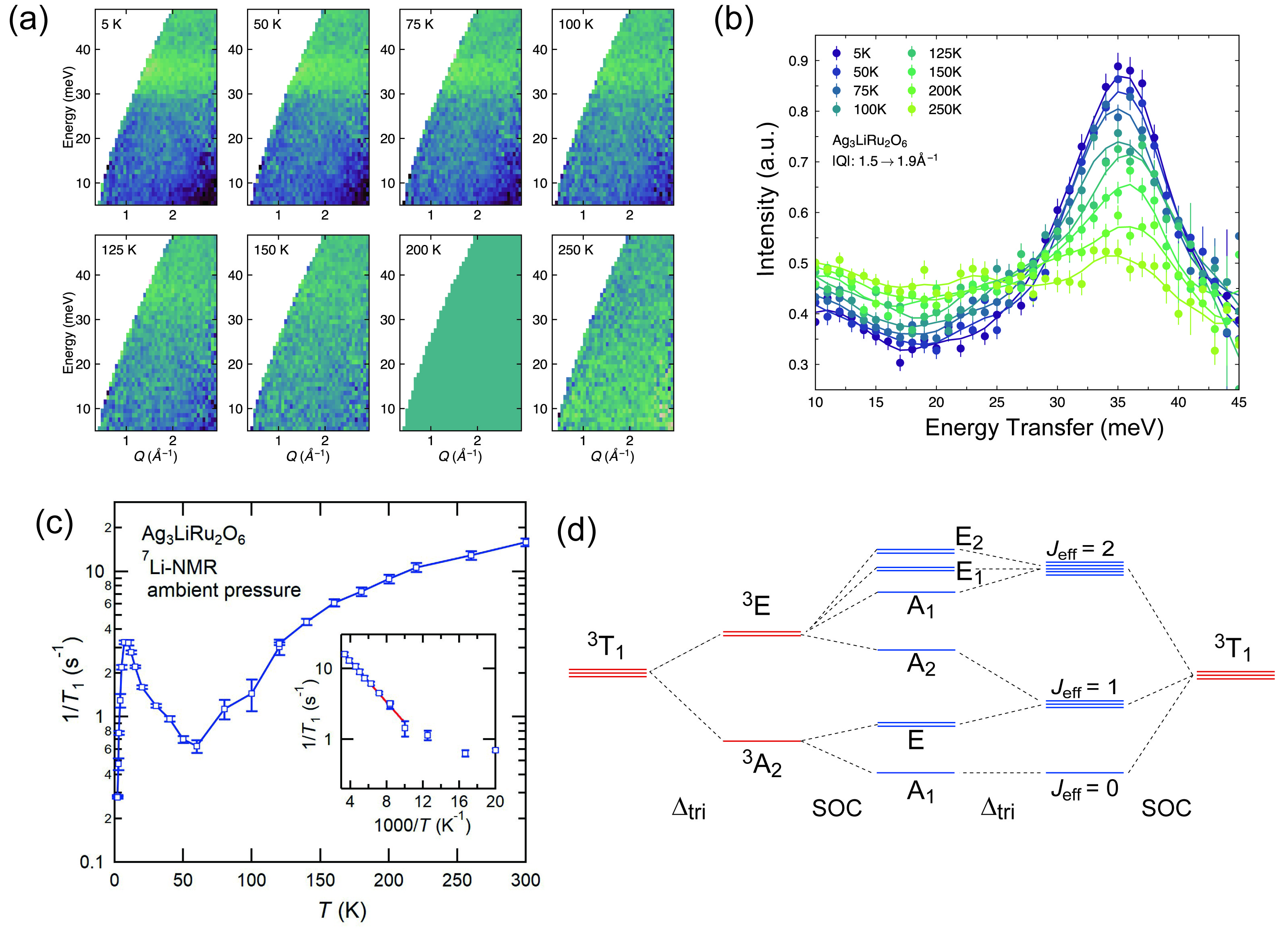}
\caption{Spin-orbit-entangled singlet state in Ag$_3$LiRu$_2$O$_6$ at ambient pressure. (a) Inelastic neutron scattering (INS) on the powder sample of Ag$_3$LiRu$_2$O$_6$. The figures show the contour plot of intensities normalized by the data at 200 K at small scattering wave vector $Q$. The raw data are shown in the Supplemental Fig. S5 \cite{SM}. (b) INS intensity as a function of energy transfer in the $\left|Q\right|$-region between 1.5 and 1.9 \AA$^{-1}$ at various temperatures. (c) Inverse of $^7$Li-NMR spin-lattice relaxation time $T_1$ as a function of temperature. $T_1$ is measured at the larger shift peak of spectra (see Fig.~\ref{fig:3}(b)). The inset shows the Arrhenius plot of the data, and the red line represents the fit between 100 and 300 K. The fit gives a gap of $\sim$30 meV. The peak of 1/$T_1$ around 10 K is likely attributed to the localized spin defects observed in $\chi(T)$ and $C(T)$. (d) Schematic energy levels of Ru$^{4+}$ $t_{2g}^4$ states obtained by the quantum chemistry calculation. The states expressed by red bar are 3-fold degenerate because of $S$ = 1, while those of blue bar represent a single state. Each state is labelled with a Mulliken symbol. The nine states of $^3$T$_1$ split by trigonal crystal field ($\Delta_{\rm tri}$) into the $^3$A$_2$ and $^3$E states. Spin-orbit coupling (SOC) further splits those states and yields the A$_1$ singlet ground state. The right-side of the diagram shows the $J_{\rm eff}$-picture where SOC first splits the $^3$T$_1$ states into $J_{\rm eff}$ = 0, 1, and 2 states and $\Delta_{\rm tri}$ acts on the $J_{\rm eff}$-states. The resultant energy diagram is identical in both representations.
}
\label{fig:2}
\end{center}
\end{figure*}

The $^7$Li-NMR spectrum on powder sample at room temperature shows a quite asymmetric line-shape, which we ascribe to a magnetic anisotropy (Fig.~\ref{fig:3}(b) and Supplemental Fig. S4 \cite{SM}). There is only one Li crystallographic site in Ag$_3$LiRu$_2$O$_6$ and inter-site mixing between Li and Ag was not identified in the analysis of powder x-ray diffraction \cite{Bette2019}. Upon cooling, the asymmetry increases but no splitting is observed down to 5 K, demonstrating the absence of magnetic ordering. In the corresponding spin-lattice relaxation rate 1/$T_1$ shown in Fig.~\ref{fig:2}(c), no divergence indicative of magnetic phase transition is observed. The broad peak at a low temperature can be ascribed to contributions from magnetic defects seen in $\chi$($T$) and $C$($T$).

The time dependence of muon asymmetry (Fig.~\ref{fig:1}(d)) shows a monotonic decrease without any oscillatory signals, again reinforcing the absence of magnetic order. The time spectra show the Gaussian shape at high temperatures, whereas they change into the exponential curve below 25 K (Supplemental Table S1 \cite{SM}). This behavior is reminiscent of slowing down of spin fluctuations. However, the relaxation at low temperatures is suppressed by the application of a small longitudinal field (LF) of 100 Gauss, as shown in the inset of Fig.~\ref{fig:1}(d). This indicates that the internal field at the muon sites is quite small and of the order of nuclear dipole fields. The slow relaxation under LF shows a stretched exponent behavior, exp\{$-(\lambda t)^{\beta}$\} with $\beta$ smaller than 1. This suggests that the dynamical relaxation is not homogeneous, likely associated with dilute defect spins in a nonmagnetic state. This is in line with the spin defect picture which has emerged from the other probes. 

\begingroup
\renewcommand{\arraystretch}{1.2}
\begin{table*}[t]
\centering
\caption{Energy levels of Ru$^{4+}$ $t_{2g}^4$ states by crystal field and spin-orbit coupling (SOC) in Ag$_3$LiRu$_2$O$_6$. The energy of multiplets is obtained from the embedded cluster complete active space self consistent-field (CASSCF(4e, 5o), denoted as CAS in the table) and multireference configuration interaction (MRCI) calculations, with and without incorporating SOC. The energy of the lowest state is set to be zero, and the energy gap from the lowest state are shown. The calculation was performed for the crystal structure at room temperature reported in Ref. [\onlinecite{Kimber2010}]. The nine states of $t_{2g}^4$ configuration ($^3$T$_1$ with $S$ = 1, $L$ = 1) are split into $^3$A$_2$ (3 states) and $^3$E (6 states) by trigonal crystal field. SOC further splits the $^3$A$_2$ ($^3$E) states into A$_1$ singlet and E doublet (A$_2$ and A$_1$ singlets, E$_1$ and E$_2$ doublets), respectively (see Fig.~\ref{fig:2}(d) as well). This gives rise to a gap of $\sim$47 meV between the ground state singlet A$_1$ and the lowest doublet E. Note that there is a small split within the E doublets because of weak non-trigonal distortion. }
\begin{tabular}{|c|c|c|c|} \hline
CAS (meV) & CAS+SOC (meV) & MRCI (meV) & MRCI+SOC (meV) \\ \hline
\multirow{2}{*}{$^3$A$_2$: 0} & A$_1$: 0 &\multirow{2}{*}{$^3$A$_2$: 0} & A$_1$: 0 \\ \cline {2-2} \cline{4-4}
      & E: 49.7, 51.3 &  &  E: 46.5, 48.1 \\ \hline 
\multirow{4}{*}{$^3$E: 107.1, 115.5}  & A$_2$ : 139.1 & \multirow{4}{*}{$^3$E: 119.7, 129.6} & A$_2$: 147.6 \\ \cline{2-2} \cline{4-4}
      &  A$_1$ : 198.5 & & A$_1$: 200.1 \\ \cline{2-2} \cline{4-4}
      &  E$_1$: 212.0, 218.8 & & E$_1$: 216.0, 224.3 \\ \cline{2-2} \cline{4-4}
      &  E$_2$: 235.1, 235.2 & & E$_2$: 241.5, 241.6 \\ \hline
\end{tabular}
\label{table1}
\end{table*}
\endgroup

\subsection{Spin-orbit-entangled singlet state at ambient pressure}

The nonmagnetic state with van Vleck-like behavior of $\chi$($T$) strongly suggests a spin-orbit-entangled singlet ground state in Ag$_3$LiRu$_2$O$_6$, which is characterized by the presence of a small excitation gap to $J_{\rm eff}$ = 1 states of the order of SOC. The presence of a magnetic excitation gap of 35 meV is indeed identified by a time-of-flight powder inelastic neutron scattering (INS) as shown in Fig.~\ref{fig:2}(a). There is an intense signal centered at around 35 meV in the region of small momentum transfer $\left|Q\right|$, and its intensity decreases quickly as $\left|Q\right|$ increases, indicating the magnetic origin of excitation. The integrated intensity in a small $Q$ region (1.5 – 1.9 \AA$^{-1}$) monotonically decreases on heating as shown in Fig.~\ref{fig:2}(b) but can be visible even at 200 K.

The presence of a magnetic excitation gap is also confirmed in the temperature dependence of spin lattice relaxation rate 1/$T_1$ from $^7$Li-NMR. As seen in Fig.~\ref{fig:2}(c), 1/$T_1$ exhibits a complex temperature dependence: on cooling from room temperature, 1/$T_1$ decreases rapidly and then shows a dip at about 50 K followed by a broad peak at around 10 K, which we attributed to the low-lying excitations associated with the defect spins. The rapid decrease of 1/$T_1$ below room temperature is incompatible with a Curie-Weiss behavior of localized moments, supporting that the increase of $\chi$($T$) around room temperature originates from the van Vleck susceptibility, and therefore should be ascribed to the magnetic excitation predominantly to $J_{\rm eff}$ = 1 states. The Arrhenius plot of 1/$T_1$ above 100 K, shown in the inset of Fig.~\ref{fig:2}(c), yields a gap of $\sim$30 meV, close to the one observed in INS. The gap between the $J_{\rm eff}$ = 0 singlet and the $J_{\rm eff}$ = 1 triplet corresponds to $\lambda_{\rm SO}$ ($\lambda_{\rm SO} = \zeta/2S$ where $\zeta$ is the single-electron SOC and $S$ = 1). The free-electron value of $\zeta$ for Ru is $\sim$140 meV, and thus the expected gap is $\sim$70 meV as observed in a $J_{\rm eff}$ = 0 Mott insulator K$_2$RuCl$_6$ \cite{Takahashi2021}. This gap value is much larger than $\sim$35 meV observed in Ag$_3$LiRu$_2$O$_6$. We argue that the trigonal distortion of the RuO$_6$ octahedra splits the $J_{\rm eff}$ = 1 triplet into a singlet and doublet and reduces the lowest excitation gap as schematically illustrated in Fig. 2(d). Essentially the same mechanism of gap reduction was discussed in Ca$_2$RuO$_4$ hosting tetragonally-distorted RuO$_6$ octahedra \cite{Jain2017}.

Embedded cluster quantum chemistry calculations of electronic structure were performed with the complete-active-space-self-consistent-field (CASSCF) and the multireference configuration-interaction (MRCI) methods and give the energy level scheme summarized in Fig.~\ref{fig:2}(d) and Table~\ref{table1}. The calculations without SOC show that the $t_{2g}^4$ multiplet of a Ru$^{4+}$ ion with $S$ = 1 and $L_{\rm eff}$ = 1 ($^3$T$_1$) splits into the lower $^3$A$_2$ and the upper $^3$E states in the presence of trigonal crystal field. The two methods yield very close results and we hereafter refer only to the one obtained by the more accurate MRCI treatment. The splitting of $^3$A$_2$ and $^3$E is $\sim$120 meV, which is reasonable in magnitude as the trigonal crystal field splitting of $t_{2g}$ multiplet in 4$d$ transition-metal oxides with an octahedral coordination. There is an additional small split of the originally-degenerate $^3$E state due to the presence of non-trigonal crystal field originating from the monoclinic distortion, which we ignore in the following discussion for simplicity. This small split can be found in the Table~\ref{table1} and indeed negligibly small as compared with the trigonal field splitting. By incorporating SOC, the $^3$A$_2$ state splits into the ground state A$_1$ singlet and the upper E doublet $\sim$47 meV above the A$_1$. The higher $^3$E state also splits by SOC into the two singlets (A$_2$ and A$_1$ at $\sim$148 meV and $\sim$200 meV above the ground state A$_1$ singlet, respectively) and two doublets (E$_1$ and E$_2$ at $\sim$220 meV and $\sim$240 meV). The singlet A$_2$ is the lowest excited state out of the $^3$E state. In the $J_{\rm eff}$-language, the ground state A$_1$ singlet originates from the $J_{\rm eff}$ = 0 state and the low-lying excited states E doublet and A$_2$ singlet from the $J_{\rm eff}$ = 1 triplet. That is, the $J_{\rm eff}$ = 1 triplet splits into the lower E doublet at $\sim$47 meV and the upper A$_2$ singlet at $\sim$148 meV with the trigonal crystal field (see the right-side of Fig.~\ref{fig:2}(d)). We note that the A$_1$ singlet ground state is no longer a pure $J_{\rm eff}$ = 0 state in the presence of trigonal distortion.

The calculated A$_1$ singlet - E doublet gap of $\sim$47 meV agrees reasonably with the excitation gap $\sim$35 meV seen in the INS measurement. We do not resolve clear dispersion in the $\sim$35 meV excitation in the INS data with $\sim$10 meV width, which may suggest a small exchange coupling and negligible dispersion of excited states as compared with the excitation gap. From these experimental and theoretical results, we conclude that the ambient pressure phase of Ag$_3$LiRu$_2$O$_6$ is $J_{\rm eff}$ = 0-derived spin-orbit-entangled singlet state, which we call $J$-singlet state hereafter.

\subsection{Pressure-induced phase transitions}

\begin{figure*}[t]
\begin{center}
\includegraphics[scale=0.84]{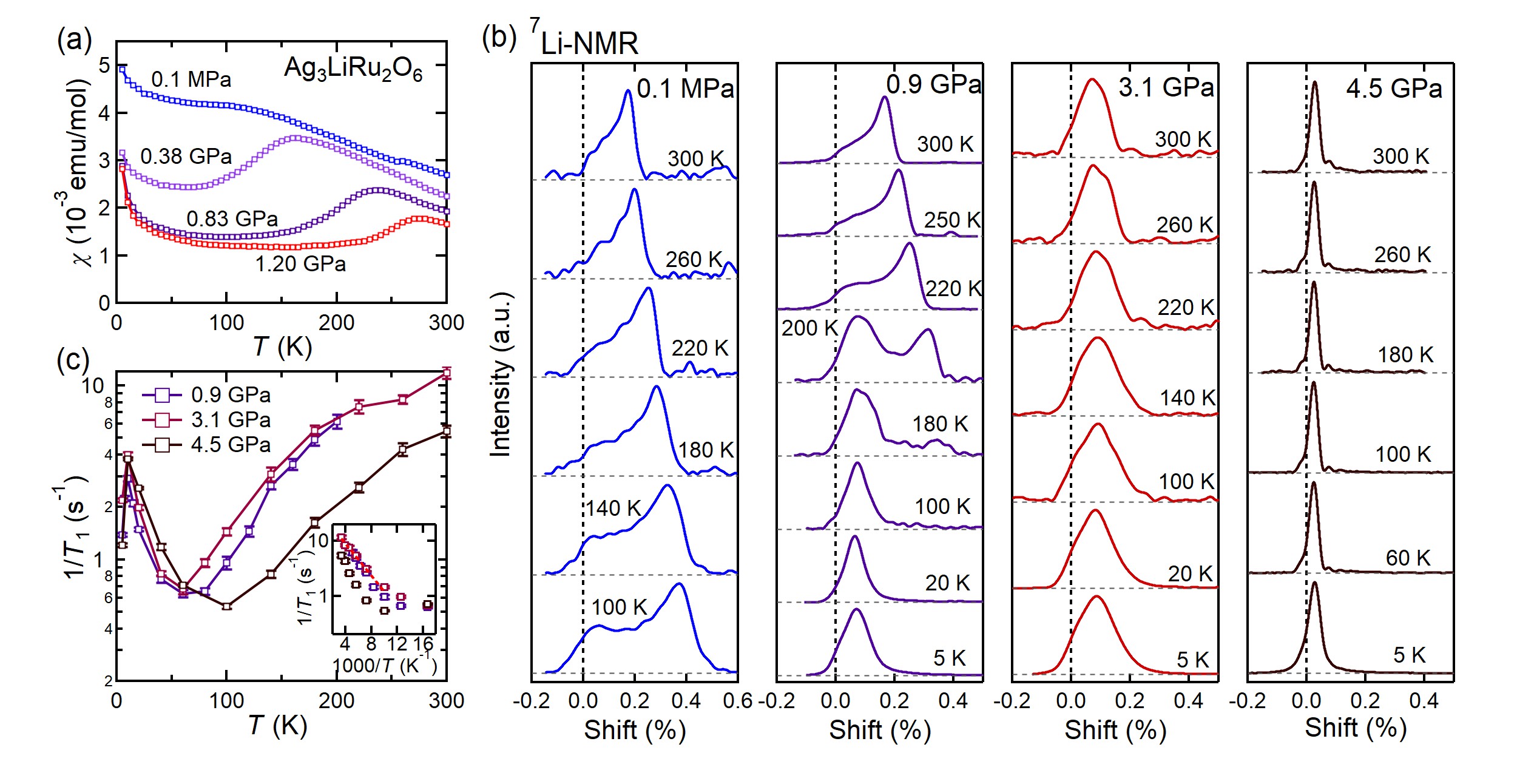}
\caption{Pressure-induced change of magnetism in Ag$_3$LiRu$_2$O$_6$. (a) Magnetic susceptibility $\chi(T)$ under pressure. (b) $^7$Li-NMR spectra at various pressures. The two-peak structure at ambient pressure likely originates from the strong magnetic anisotropy (see Supplementary Fig. S4 \cite{SM}). At 0.9 GPa, a change of spectrum from an asymmetric shape to a symmetric and lower shift peak was seen on cooling below 200 K. The spectra at 0.9 and 3.1 GPa at low temperatures remain sharp, pointing to a nonmagnetic nature of the intermediate phase. At 4.5 GPa, the spectra become much sharper than the lower pressure data. (c) Inverse of spin-lattice relaxation time $T_1$ under pressure obtained from $^7$Li-NMR. The 1/$T_1$ at 0.9 GPa was obtained at the smaller shift peak below 200 K. The inset shows the Arrhenius plot of 1/$T_1$ where the similar gapped behavior as in the ambient pressure phase is seen. The dotted line shows a fit for the 3.1 GPa data between 300 K and 100 K. The estimated gap sizes at 0.9 and 3.1 GPa are nearly same ($\sim$30 meV), while the one at 4.5 GPa is larger ($\sim$48 meV), implying a different origin of excitation gap.
}
\label{fig:3}
\end{center}
\end{figure*}

To explore the possible excitonic magnetism and other exotic states out of the $J$-singlet state, we attempted to enhance the exchange interactions by applying pressure. At ambient pressure, $\chi$($T$) shows a van Vleck-like behavior as discussed above. With the application of pressure, $\chi$($T$) shows a drastic change as shown in Fig.~\ref{fig:3}(a). At 0.38 GPa, $\chi$($T$) displays a shoulder-like anomaly around 150 K and shows an almost temperature-independent behavior at lower temperatures, indicative of a pressure-induced transition. The transition temperature, defined as the temperature for $\chi$($T$) anomaly, increases rapidly to almost room temperature with increasing pressure above 1 GPa. The temperature-dependent resistivity does not change appreciably under pressure with no clear anomaly at the transition (Supplemental Fig. S6 \cite{SM}).

The pressure-induced phase, which we call an intermediate phase because of the presence of another phase at a higher pressure, is nonmagnetic and not an excitonic magnet with closed gap. $^7$Li-NMR spectra under pressure in Fig.~\ref{fig:3}(b) indicate the absence of magnetic order in the intermediate phase. At $P$ = 0.9 GPa, the asymmetric spectrum around room temperature does not change substantially from that at ambient pressure. Below 200 K where the $\chi$($T$) anomaly is seen at a similar pressure (0.83 GPa in Fig.~\ref{fig:3}(a)), however, the asymmetric peak fades out and is replaced by one symmetric peak with a small shift at low temperatures, supportive of the occurrence of phase transition around 200 K. The spectrum at around 200 K comprises the superposition of the high-temperature asymmetric peak and the low-temperature symmetric peak, which indicates the two-phase coexistence and hence the first order nature of transition. The symmetric peak in the intermediate phase remains relatively sharp down to the lowest temperature, excluding a magnetic order. The absence of magnetic order is also corroborated by the ZF-$\mu$SR measurement performed in the intermediate phase down to 2 K at 0.64 GPa, where no signature of muon precession found (Supplemental Fig. S7 \cite{SM}). At 3.1 GPa, the NMR spectra for the intermediate phase persists up to room temperature, meaning that the phase transition temperature if any is above room temperature.

Interestingly, the excitation gap estimated from 1/$T_1$ in the intermediate phase is $\sim$30 meV, not appreciably different from that of $J$-singlet phase at ambient pressure, as estimated from the Arrhenius plot at 0.9 GPa between 200 and 100 K and at 3.1 GPa in the inset of Fig.~\ref{fig:3}(c). The nature of the intermediate phase should be closely related to that of the $J$-singlet phase.

By further increasing pressure to 4.5 GPa, the $^7$Li-NMR spectrum shows a substantial change, indicative of another pressure-induced phase transition. The peak becomes much sharper compared with those at lower pressures and the peak position, i.e. Knight shift, is close to zero and independent of temperature. Such a sharp and zero-shift peak has been observed in the dimerized state of Li$_2$RuO$_3$ \cite{Arapova2017}, owing to the almost complete absence of internal magnetic fields from electron spins. This suggests the formation of strong dimers up to room temperature at 4.5 GPa. In accord with this, 1/$T_1$ above 100 K is strongly suppressed compared with those of the ambient-pressure and the intermediate phases and shows the enhanced excitation gap of $\sim$48 meV. This indicates that another pressure-induced phase with a distinct excitation gap, which we call high-pressure phase, is realized around 4.5 GPa.

\begin{figure}[t]
\begin{center}
\includegraphics[scale=0.16]{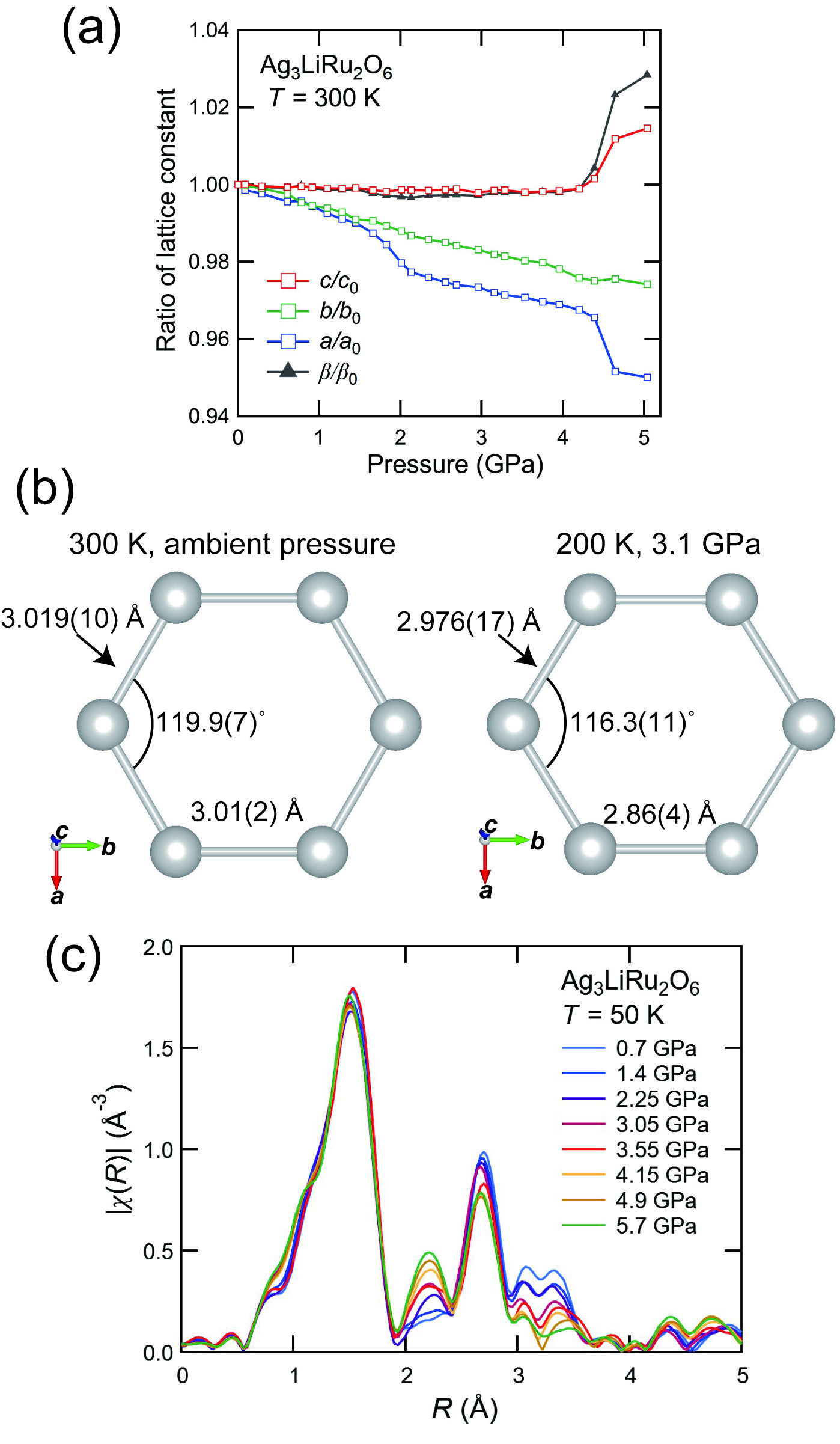}
\caption{Pressure-induced structural transitions in Ag$_3$LiRu$_2$O$_6$. (a) Pressure-dependent lattice parameters at room temperature obtained from neutron diffraction. The values of lattice constants are evaluated using the monoclinic unit cell of the ambient pressure phase (space group $C2/m$) and normalized by those at ambient pressure. (b) Ru honeycomb lattices at ambient pressure (left) and in the intermediate phase (right). The structure at ambient pressure is taken from Ref. [\onlinecite{Bette2019}], and the one in the intermediate phase was obtained from the Rietveld refinement of neutron diffraction at 200 K and at 3.1 GPa. (c) The magnitude of the complex Fourier transform (FT) of XAFS data for various pressures at $T$ = 50 K. Note that the FT is not corrected for scattering phase shifts, and the $R$ values do not exactly correspond to the bond lengths. }
\label{fig:4}
\end{center}
\end{figure}

Neutron diffraction under pressure reveals that the two pressure-induced phase transitions, identified in $\chi$($T$) and NMR, are accompanied by a structural change. The lattice parameters at room temperature under pressure were refined using the monoclinic unit cell of the ambient pressure phase, which is displayed in Fig.~\ref{fig:4}(a) as a function of pressure. The $a$-axis and the $b$-axis are in the honeycomb plane, along one of the Ru zigzag chains of honeycomb lattice and parallel to the bridging Ru-Ru bonds respectively, as depicted in Fig.~\ref{fig:4}(b). The $c$-axis defines the out-of-plane direction. The in-plane lattice constants, $a$ and $b$, decrease with pressure, while $c$ and the monoclinic angle $\beta$ remain almost constant up to $\sim$4 GPa. At around 1.8 GPa, $a$ shows a discontinuous drop, indicating the presence of a first-order structural transition. By increasing pressure further, another sudden change of lattice parameters is observed at $\sim$4.5 GPa, where $a$ shrinks further and $c$ and $\beta$ increase. Upon cooling, the critical pressures for the two structural transitions are reduced (Supplemental Fig. S8 \cite{SM}). The critical pressures for the first structural transition agree reasonably with those determined from $\chi$($T$) as depicted in the phase diagram shown in Fig.~\ref{fig:5}, indicating that the transition to the intermediate phase is accompanied by the first structural transition. The second transition represents the sharpening of $^7$Li-NMR spectra observed at 4.5 GPa, namely the transition from the intermediate phase to the high-pressure phase. 

The crystal structure of the intermediate phase was refined reasonably by the same structural model with that at ambient pressure (Supplemental Table S2 \cite{SM}). While there is no pronounced change in the distortion of RuO$_6$ octahedra (Supplemental Fig. S9 \cite{SM}), the honeycomb lattice of Ru atoms is weakly squashed along the $a$-axis. The Ru honeycomb lattice refined at 200 K and 3.1 GPa is illustrated in Fig.~\ref{fig:4}(b), together with that at ambient pressure \cite{Bette2019}. The angle of Ru-Ru bonds along the zigzag chains is decreased to $\sim$116$^\circ$ from almost 120$^\circ$ at ambient pressure, which results in the shortening of the bridging Ru-Ru bonds along the $b$-axis by about 4\% compared with the other bonds in the zigzag chain. Hence, this intermediate phase is characterized by the presence of weak Ru-Ru dimers along the $b$-axis.

The drastic sharpening of NMR linewidth at 4.5 GPa suggests the change of electronic ground state through the second structural transition. The refinement of neutron diffraction pattern in the high-pressure phase was not successful with using the structural model of the intermediate phase, implying a distinct distortion of the Ru-honeycomb lattice. To investigate the distortion, x-ray absorption fine structures (XAFS) spectra at Ru $K$-edge were collected under pressure. Figure~\ref{fig:4}(c) shows the Fourier-transform (FT) magnitude of complex XAFS function at 50 K, related to the partial radial distribution function of Ru atoms. The peak at $R$ $\sim$ 1.5 \AA~ represents the Ru-O bonds, while that at $\sim$2.7 \AA~ corresponds to the Ru-Ru distance. Note that FT data shown in Fig.~\ref{fig:4}(c) is not corrected for photoelectron scattering phase shifts in the central and neighboring atoms and thus the peaks in the FT are shifted to smaller distances. By inspecting the peak intensities at $\sim$2.7 \AA, a clear difference is seen between the pressures below 3.05 GPa and above 4.15 GPa. The structural transition appears to occur at around 3.5 GPa which agrees with the phase boundary estimated from the neutron diffraction results (Fig.~\ref{fig:5}). In the high-pressure phase, the growth of the peak is seen at $R$ $\sim$ 2.2 \AA. This peak may indicate the shortening of Ru-Ru distance. We analyzed the data at 5.7 GPa with the strong dimer model similar to the structure of Li$_2$RuO$_3$, where a Ru-hexagon is composed of two short bonds and four long bonds. The refinement of FT data gives the Ru-Ru distances of 2.51(2) \AA~ and 3.06(2) \AA~ for the short and long bonds, respectively (Supplemental Fig. S10 \cite{SM}). The difference of bond lengths is $\sim$19\% which is close to that in Li$_2$RuO$_3$ ($\sim$17\%), indicating the formation of strong Ru-Ru dimers induced by pressure. We note that the peak at $\sim$2.2 \AA~ starts to grow already at $\sim$2.3 GPa, i.e. in the intermediate phase. This may indicate that local dimerization takes place partially in the intermediate phase or that fluctuating dimers are present as a precursor of strong dimers as discussed in the high-temperature phase of Li$_2$RuO$_3$ \cite{Kimber2014}.

\section{Discussion}

The honeycomb ruthenate Ag$_3$LiRu$_2$O$_6$ was found to host a spin-orbit-entangled $J$-singlet state, proximate to $J_{\rm eff}$ = 0 state, at ambient pressure. This state is realized by suppressing the Ru-Ru dimerization that takes place in Li$_2$RuO$_3$. The honeycomb lattice of spin-orbit-entangled singlet is expected to display frustrated excitonic magnetism such as a spin-nematic state and the bosonic Kitaev liquid. However, the $J$-singlet state remains intact down to the lowest temperature measured at ambient pressure without any discernible changes in the magnetic excitation spectra. This is likely because the exchange interactions via the upper doublet, derived from the $J_{\rm eff}$ = 1 triplet, are not strong enough. In order to realize the excitonic magnetism, enhancement of exchange interactions and/or the reduction of singlet-doublet gap with stronger trigonal crystal field would be required.

Upon the application of pressure, the Ru honeycomb lattice of Ag$_3$LiRu$_2$O$_6$ first shrinks almost isotropically (Fig.~\ref{fig:4}(a)). In tandem with this, magnetic susceptibility at room temperature decreases monotonically up to $\sim1.2$ GPa (Fig.~\ref{fig:3}(a)), implying an enhancement of antiferromagnetic interactions. 
By further increasing pressure, instead of developing excitonic magnetism, Ag$_3$LiRu$_2$O$_6$ exhibits successive phase transitions to other nonmagnetic phases. The first transition to the intermediate phase at $\sim$1.8 GPa at room temperature is characterized by the modest squashing of Ru honeycomb lattice along the $a$-axis and the formation of weak Ru-Ru dimers along the $b$-axis. The intermediate phase was found to be nonmagnetic by the $^7$Li-NMR and $\mu$SR measurements. 

The second transition to the high-pressure phase appears at around 4.5 GPa at room temperature, accompanied by the formation of strong Ru-Ru dimers with very short bond-length. This dimer phase is reminiscent of the low temperature phase of Li$_2$RuO$_3$ where the large bonding-antibonding split of $d$-electron molecular orbitals (MO) stabilizes the strong Ru-Ru dimers \cite{Miura2007,Miura2009}. The spin and orbital degrees of freedom are fully quenched in the MOs, leading to a nonmagnetic ground state. The sharp and almost zero shift $^7$Li-NMR peak observed at 4.5 GPa indicates that such a MO state is also present in the high-pressure phase of Ag$_3$LiRu$_2$O$_6$. We call the high-pressure phase of Ag$_3$LiRu$_2$O$_6$ the MO-dimer state. A pressure-induced strong dimerization similar to Ag$_3$LiRu$_2$O$_6$ was also observed in the honeycomb iridates $\alpha, \beta$-Li$_2$IrO$_3$ \cite{Hermann2018, Takayama2019} and $\alpha$-RuCl$_3$ \cite{Biesner2018}, indicating that the competition between spin-orbital entanglement and a MO state is common in spin-orbit-coupled honeycomb compounds.

\begin{figure}[t]
\begin{center}
\includegraphics[scale=1.0]{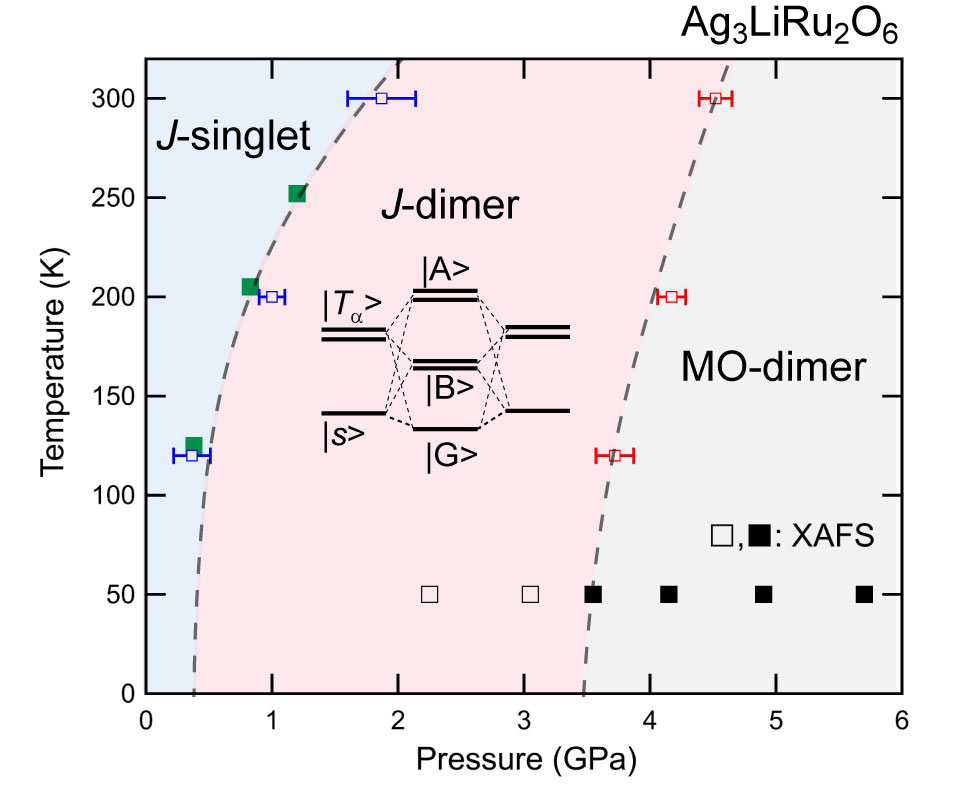}
\caption{Pressure-temperature phase diagram of Ag$_3$LiRu$_2$O$_6$. The blue and red open squares represent the phase boundaries determined by the neutron diffraction, whereas the green squares show the transition temperatures determined by d$\chi(T)$/d$T$. The filled (open) black squares indicate the points where the strong dimerization of Ru atoms was found in the XAFS spectra in the high-pressure (intermediate) phases, respectively. The inset depicts the $J$-dimer model of the intermediate phase. On each site, we consider a singlet $\ket{s}$ and upper doublet $\ket{T_{\alpha}}$ ($\alpha = x, y$) which is split from the $J_{\rm eff}$ = 1 triplet by trigonal crystal field. These states form the ground state $\ket{\rm G}$, ``bonding'' states $\ket{\rm B}$ and ``antibonding'' states $\ket{\rm A}$ by exchange interactions. In the full $J$-dimer model including the upper singlet $\ket{T_{z}}$, there are 16 states per dimer in total, but we show here only the lower 5 states for brevity. See Supplemental Material \cite{SM} for more discussions about the $J$-dimer model.
}
\label{fig:5}
\end{center}
\end{figure}

The nature of the intermediate phase is distinct from the high-pressure phase with the MO-dimers. The magnetic susceptibility $\chi$($T$) of the intermediate phase is approximately 1.5 $\times$ 10$^{-3}$ emu/Ru-mol from the data at 1.20 GPa, which is much larger than that of MO phase of Li$_2$RuO$_3$ ($\sim$0.3 $\times$ 10$^{-3}$ emu/Ru-mol \cite{Miura2007}). While the intermediate phase is nonmagnetic, the $^7$Li-NMR spectra are broader than those in the high-pressure phase, which may indicate non-negligible magnetic anisotropy and that the spin-orbit coupling is not quenched. The DFT calculations on this phase point to a metallic state despite the insulating behavior of the resistivity, suggesting a Mott insulating state (Supplemental Fig. S12 \cite{SM}). This contrasts with the systems with strong dimers where DFT calculations yield a band insulating ground state associated with molecular orbital formation \cite{Miura2009,Takayama2019,Yun2019}. These facts suggest that the weak dimers of the intermediate phase do not involve the formation of molecular orbitals. The excitation gap of the intermediate phase, which is close to that of the ambient pressure phase estimated from the NMR 1/$T_1$ (Fig.~\ref{fig:3}(c)), implies that the intermediate phase maintains a character of a spin-orbit-entangled singlet state where the excitation gap is determined by the strength of SOC but is stabilized by the formation of weak dimers. 

The basic electronic structure of the intermediate phase, associated with the weak dimer of the $J$-singlet states, can be understood as schematically illustrated in the inset of Fig.~\ref{fig:5}. The short Ru-Ru distance should enhance the exchange interactions on the weak dimer bonds, which leads to the split of excited triplet states into the states with ``bonding'' and ``antibonding'' characters (Supplemental Fig. S13 \cite{SM}) and more importantly allows the hybridization between the ground state singlet and the excited triplet states. The ground state $\ket{\rm G}$ consists of a pair of singlet $\ket{ss}$ with a small admixture of triplet pairs $\ket{T_{\alpha}T_{\alpha}}$ where $T_{\alpha}$ denotes the triplet with three different components $\alpha = x$, $y$, and $z$ \cite{Khaliullin2013,Chaloupka2019}. The bonding $\ket{\rm B}$ (antibonding $\ket{\rm A}$) states comprise the singlet and triplet pairs, $\ket{sT_{\alpha}} - \ket{T_{\alpha}s}$ ($\ket{sT_{\alpha}} + \ket{T_{\alpha}s}$), respectively (Supplemental Table S3 \cite{SM}). In Ag$_3$LiRu$_2$O$_6$, the $J_{\rm eff}$ = 1 triplet splits into the lower doublet and the upper singlet by the trigonal crystal field. Nevertheless, the lower doublet similarly forms the ``bonding'' and ``antibonding'' states as depicted in the inset of Fig.~\ref{fig:5} and Supplemental Fig. S14 \cite{SM}. We call this intermediate phase with weak dimers the $J$-dimer state. While both the $J$-singlet and the $J$-dimer states are spin-orbital singlets, the lattice distortion of the $J$-dimer state renders the admixture of low-lying excited states derived from $J_{\rm eff} = 1$ triplet into the ground state singlet and lowers its energy, which can be viewed as a pseudo-Jahn-Teller (JT) effect \cite{Liu2019}. Such weak-dimer distortion has not been observed in $d^5$ honeycomb-based iridates and $\alpha$-RuCl$_3$ under pressure, particularly in Ag$_3$LiIr$_2$O$_6$ (Supplemental Fig. S11 \cite{SM}) and Cu$_2$IrO$_3$ \cite{Fabbris2021} comprising similar interlayer dumbbell bonds with those of Ag$_3$LiRu$_2$O$_6$. From this, we infer that the weak-dimer distortion is induced by the pseudo-JT effect with the presence of low-lying spin-orbital excited states rather than structural instability and thus is unique to $d^4$ honeycomb systems.

The apparent decrease of low-temperature $\chi$($T$) from the ambient-pressure $J$-singlet state to the intermediate-pressure $J$-dimer phase, if it is dominated by van Vleck process, indicates the increase of the excitation gap, in contrast to the robust gap magnitude estimated from 1/$T_1$. The $J$-dimer picture described above can reasonably explain the contrasted behavior of $\chi$($T$) and 1/$T_1$. In the $J$-dimer phase, there are two low-energy magnetic excitations, from $\ket{\rm G}$ to $\ket{\rm B}$ and to $\ket{\rm A}$. Under a magnetic field, there is a finite mixing between $\ket{\rm G}$ and $\ket{\rm A}$ but ``not'' between $\ket{\rm G}$ and $\ket{\rm B}$ (Supplemental Table S4~\cite{SM}). The mixing between $\ket{\rm G}$ and $\ket{\rm A}$, with a larger gap than that between $\ket{\rm G}$ and $\ket{\rm B}$, therefore determines the van Vleck magnetic susceptibility. On the other hand, NMR 1/$T_1$, which is in proportion to the $q$-integrated imaginary part of dynamical susceptibility, should capture both $\ket{\rm B}$ and $\ket{\rm A}$ excitations. The magnitude of excitation gap estimated from 1/$T_1$ in the intermediate phase should reflect the average of the excitations to $\ket{\rm B}$ and to $\ket{\rm A}$ and thus be not so different from that at ambient pressure. In fact, the energy level splitting of weak dimer depends on the dominant exchange interactions: the Kitaev-type exchange acts on only one component of triplet ($\ket{T_z}$ along the $z$-bond) while all components are involved in the Heisenberg-type coupling (Supplemental Fig. S13 \cite{SM}). The measurement of a detailed excitation spectrum in the intermediate phase should provide further support for the $J$-dimer state and information about the dominant exchange interaction.

\section{Conclusion}

A honeycomb lattice of spin-orbit-entangled singlets was identified in Ag$_3$LiRu$_2$O$_6$ with 4$d^4$ Ru$^{4+}$ ions, indicating that honeycomb ruthenates are the promising candidate system to realize unconventional excitonic magnetic phases. Ag$_3$LiRu$_2$O$_6$ is a nonmagnetic Mott insulator with $J_{\rm eff}$ = 0-derived $J$-singlet state at ambient pressure due to small exchange interactions through the excited $J_{\rm eff}$ = 1-derived states. By application of pressure, Ag$_3$LiRu$_2$O$_6$ displays successive phase transitions to other nonmagnetic phases instead of developing excitonic magnetism. While the high-pressure MO phase is analogous to those identified in other honeycomb-based materials, the intermediate phase with a spin-orbit $J$-dimer state is unique to this honeycomb ruthenate and has not been predicted in theory. We argue that the $J$-dimer state is induced by a pseudo-JT effect associated with low-lying spin-orbital excitations. The pseudo-JT effect, which is likely inherent in spin-orbit-entangled $d^4$ compounds, may compete with the development of frustrated excitonic magnetism. Therefore the role of the lattice degree of freedom should be explicitly taken into account for further materials design. We believe this finding will open up a pathway for rich physics of $J_{\rm eff}$ = 0-based honeycomb systems as have been established in the celebrated $J_{\rm eff}$ = 1/2 Kitaev materials \cite{Jackeli2009,Takagi2019}.

\section*{ACKNOWLEDGMENTS}

We are grateful to G. Khaliullin, J. Chaloupka and G. Jackeli for invaluable discussions. We acknowledge the provision of beamtimes on PEARL (Proposal No. RB1920241 \cite{PEARL}) and MAPS (Proposal No. RB1820520 \cite{MAPS}) to Science \& Technology Facilities Council (STFC). We thank the RIKEN-RAL muon facilities for the allocation of beamtime on ARGUS at ISIS Neutron and Muon Source (Proposal No. RB1970001 \cite{muon}). We thank the National Synchrotron Radiation Research Center (NSRRC) and the Japan Synchrotron Radiation Research Institute (JASRI) for the allocation of beamtime for high-pressure x-ray diffraction measurements at BL12B2 of SPring-8 (Proposal No. 2018B4139). Work at Argonne is supported by the US Department of Energy, Office of Science, Office of Basic Energy Sciences, under Contract No. DE-AC- 02-06CH11357. We thank PRIUS for providing nanopolycrystalline diamond anvils. W. B. acknowledges partial support by the Consortium for Materials Properties Research in Earth Sciences (COMPRES). This work is partly supported by Alexander von Humboldt Foundation.


\end{document}